\documentclass[11pt,a4paper]{JHEP3}

\usepackage{amsmath,amssymb}

\def\ap#1{\alpha'{}^{\,#1}}
\def\a{\alpha}

\def\G{\Gamma}

\def\l{\lambda}

\def\m{\mu}
\def\n{\nu}

\def\r{\rho}

\def\w{\omega}
\def\W{\Omega}

\def\be{\begin{equation}}
\def\ee{\end{equation}}
\def\bea{\begin{eqnarray}}
\def\eea{\end{eqnarray}}

\def\({\mbox{$(\!($}}
\def\){\mbox{$)\!)$}}

\title{$ \ap{}$-Corrections to Heterotic Superstring Effective Action Revisited}
\author{W.~A. Chemissany, Mees de Roo\\
Centre for Theoretical Physics, University of Groningen,\\
Nijenborgh 4, 9747 AG Groningen, The Netherlands\\
E-mail: \email{W.Chemissany@rug.nl, M.de.Roo@rug.nl }}
\author{Sudhakar Panda\\
Harish-Chandra Research Institute \\
Chatnag Road, Jhusi, Allahabad 211019, India\\
E-mail: \email{panda@mri.ernet.in}}

\preprint{
UG-07-04}


\abstract{In this letter we establish that the
supersymmetric $R^2$ effective action for the heterotic string,
obtained from the supersymmetrisation of the 
Lorentz Chern-Simons term, is to order $\alpha'$ equivalent modulo 
field redefinitions to heterotic string effective actions computed 
by different methods.}

\begin{document}

\section{Introduction\label{Intro}}

The possibility to compare calculations of the entropy of certain black holes 
by microscopic string methods and by direct
methods of general relativity\footnote{For an extensive introduction
to this field see \cite{Mohaupt:2000mj}.}
has caused renewed interest in the structure of higher derivative 
contributions to the string effective action. 
In this paper we clarify the relation between two formulations
of the order $\ap{}$ heterotic string effective action. One formulation
follows from string amplitudes calculations
\cite{Gross:1986mw,Metsaev:1987zx}
and from the requirement of conformal symmetry of the corresponding sigma model
to the appropriate order \cite{Metsaev:1987zx,Hull:1987pc}, the other 
formulation \cite{Bergshoeff:1988nn,Bergshoeff:1989de} is based on the 
supersymmetrisation of Lorentz-Chern-Simons (LCS) forms. 
Our interest in the relation between these results was triggered 
by a remark in a recent paper of Sahoo and Sen \cite{Sahoo:2006rp}. 
In that paper the entropy of a supersymmetric black hole was obtained 
using the method of \cite{Sen:2005wa}, with \cite{Metsaev:1987zx} 
for the derivative corrections to the action. 
The result was found to agree with that 
obtained by several other methods, which was taken by \cite{Sahoo:2006rp}
as an indirect indication that the bosonic expression for the order 
$\ap{}$ corrections given in \cite{Metsaev:1987zx} must be part of a 
supersymmetric invariant. 

The result of \cite{Bergshoeff:1988nn} is supersymmetric to
order $\ap{}$, in \cite{Bergshoeff:1989de} results to order $\ap{2}$
and $\ap{3}$ are obtained as well. We will show in this paper that
to order $\ap{}$ \cite{Bergshoeff:1988nn} agrees
with \cite{Metsaev:1987zx}, proving directly that the action
of \cite{Metsaev:1987zx}
is indeed part of a supersymmetric invariant. The field redefinitions
required to establish this correspondence generate additional
terms at higher orders in $\ap{}$.

In Section \ref{Equiv} we establish the equivalence between the 
two effective actions. The necessary background material and conventions 
can be found in the Appendices. We discuss terms of
order $\ap{2}$ and $\ap{3}$ in Section \ref{High}. Conclusions are in
Section \ref{Conc}.

\section{The heterotic string effective action\label{Equiv}}

The heterotic string effective action to order $\ap{}$, as given in 
\cite{Metsaev:1987zx}, reads
\bea
  {\cal L}_{\rm MT} &=&  -\frac{2}{\kappa^2} e\,e^{-2\Phi}\bigg(R(\G) -
   \tfrac{1}{12} {H}_{\m\n\r}{H}^{\m\n\r} +
         4 \partial_\m\Phi\partial^\m\Phi
\label{L0MT}
\\
&&\qquad
    +\tfrac{1}{8}\ap{}
        \big\{ R_{\m\n}{}_{ab}(\G)R^{\m\n\,ab}(\G)
             -\tfrac{1}{2} R_{\m\n\,ab}(\G) H^{\m\n\,c} H^{abc}
\nonumber\\
  && \qquad
    -\tfrac{1}{8} (H^2)_{ab} (H^2)^{ab} 
  +\tfrac{1}{24}  H^4
             \big\} \bigg)\,.
\label{L1MT}
\eea 
Here
\bea
  && H_{\m\n\r}= 3\partial_{[\m}B_{\n\r]}\,,
\nonumber\\
  && H^2 = H_{abc}H^{abc}\,,\quad (H^2)_{ab} = H_{acd}H_b{}^{cd}\,,\quad H^4 = H^{abc}H_a{}^{df}H_b{}^{ef}H_c{}^{de}
   \,,
\eea
normalisations are as in \cite{Metsaev:1987zx}. 

On the other hand there is the result of supersymmetrising the LCS form of 
\cite{Bergshoeff:1988nn,Bergshoeff:1989de}. 
The bosonic terms\footnote{Throughout this paper we will only 
discuss the bosonic contributions to the effective action. Fermionic
contributions can be found in \cite{Bergshoeff:1989de}.} take on the form
\bea
   {\cal L}_{\rm BR} &=& \tfrac{1}{2} e\,e^{-2\Phi}\big[\big\{ - R(\w) -
   \tfrac{1}{12} \widetilde{H}_{\m\n\r} \widetilde{H}^{\m\n\r} +
         4 \partial_\m\Phi\partial^\m\Phi\big\}
\label{L0BR}
\\   
   &&\qquad  -\tfrac{1}{2} \alpha 
             R_{\m\n}{}_{ab}(\W_-)R^{\m\n\,ab}(\W_-) \big]\,.
\label{L1BR}
\eea
With respect to \cite{Bergshoeff:1989de} we have redefined the
dilaton and the normalisation of $B_{\m\n}$ 
(see Appendix \ref{LRsection}). In (\ref{L0BR})
$\tilde H$ contains the LCS term with $H$-torsion:
\bea
     \widetilde{H}_{\m\n\r}&=& H_{\m\n\r}
  - 6 \alpha\,{\cal O}_{3\,\m\n\r}(\W_-)\,,
\label{Htilde}
\\
  {\cal O}_{3\,\m\n\r}(\W_-)&=&
      \W_{-[\m}{}^{ab}\partial_\n\W_{-\r]}{}^{ab}
    -\tfrac{2}{3}\W_{-[\m}{}^{ab}\W_{-\n}{}^{ac}\W_{-\r]}{}^{cb}\,,
\label{LCS}
\\
  \W_{-\m}{}^{ab} &=&
       \w_{\m}{}^{ab}-\tfrac{1}{2}\widetilde{H}_\m{}^{ab}\,.
\label{Om}
\eea
The coefficient $\alpha$ is proportional to $\alpha'$, note that
the relative normalisation between the LCS term and the $R^2$ 
action is fixed.

To establish the equivalence between (\ref{L0MT},\ \ref{L1MT}) and
(\ref{L0BR},\ \ref{L1BR}) we expand $R(\W_-)$ in (\ref{L1BR}),
perform the required field redefinitions and fix the normalisations.

To start with, we have 
\be
   R_{\m\n}{}^{ab}(\W_-)=R_{\m\n}{}^{ab}(\w)
   - \tfrac{1}{2}\big({\cal D}_\m \widetilde{H}_\n{}^{ab}-
                      {\cal D}_\n \widetilde{H}_\m{}^{ab}\big)
  -\tfrac{1}{8}\big(
     \widetilde{H}_\m{}^{ac}\widetilde{H}_\n{}^{cb} - 
     \widetilde{H}_\n{}^{ac}\widetilde{H}_\m{}^{cb}
              \big)\,,
\label{ROmega} 
\ee 
where the derivatives ${\cal D}$ are covariant with respect to local 
Lorentz transformations. Clearly the substitution of (\ref{ROmega}) in
(\ref{L1BR})  gives terms similar to those in (\ref{L1MT}), 
additional terms come from expanding $\widetilde{H}$ 
(see Appendix \ref{Expand}) in 
(\ref{L0BR}). The effect of these substitutions is, to order $\alpha$: 
\bea
 {\cal L}_{\rm BR}&=&
 \tfrac{1}{2}e\,e^{-2\Phi}\big[ -R(\w) -
   \tfrac{1}{12} \bar{H}_{\m\n\r} \bar{H}^{\m\n\r} +
         4 \partial_\m\Phi\partial^\m\Phi 
\nonumber\\
&& + \alpha\,\big\{
   \tfrac{1}{2} H^{\m\n\r}\partial_{\m}(\w_\n{}^{ab}{H}_{\r}{}^{ab})
  -\tfrac{1}{2} R_{\m\n}{}^{ab}(\w)\, {H}_{\r}{}^{ab}\, H^{\m\n\r}
+  \tfrac{1}{4} H^{\m\n\r}{H}_{\m}{}^{ab}
                {\cal D}_\n {H}_{\r}{}^{ab}
  \nonumber\\
&&\qquad
 - \tfrac{1}{12} H^4 \big\}
\\ 
   &&-\tfrac{1}{2}\,\alpha\,
      \big\{
     R_{\m\n}{}^{ab}(\w)R^{\m\n}{}^{ab}(\w)
\label{term1}\\
  &&\qquad -2 R^{\m\n}{}^{ab}(\w) {\cal D}_\m H_{\n ab}
\label{term2}\\
  &&\qquad + \tfrac{1}{2}
  \big({\cal D}_\m H_\n{}^{ab}-
       {\cal D}_\n H_\m{}^{ab}\big)
     {\cal D}^\m H^{\n ab}
\label{term3}\\
 &&\qquad - R_{\m\n}{}^{ab}(\w) H^{\m ac}H^{\n cb}
\label{term4}\\
 &&\qquad + \tfrac{1}{2}
  \big({\cal D}_\m H_\n{}^{ab}-
       {\cal D}_\n H_\m{}^{ab}\big)H^{\m ac}H^{\n cb}
\label{term5}\\
 &&\qquad +\tfrac{1}{8}\big( (H^2)_{ab}(H^2)^{ab} - H^4 \big)
    \big\}\big]\,.
\label{term6} 
\eea
Here $\bar{H}$ contains the LCS term without $H$-torsion:
\be
   \bar{H}_{\m\n\r} = H_{\m\n\r}
      -6\alpha \,{\cal O}_{3\,\m\n\r}(w)\,. 
\label{Hbar}
\ee
We now rewrite the terms (\ref{term1}-\ref{term6}) in ${\cal L}_{{\rm BR}}$,
see Appendix \ref{termLR2} for details.
The result, keeping only contributions to order $\alpha$, is
\bea
 {\cal L}_{\rm BR}&=&
 \tfrac{1}{2}e\,e^{-2\Phi}\big[ -R(\w) -
   \tfrac{1}{12} \bar{H}_{\m\n\r} \bar{H}^{\m\n\r} +
         4 \partial_\m\Phi\partial^\m\Phi 
\nonumber\\
&&\qquad  -\tfrac{1}{2}\alpha\,\big\{  
     R_{\m\n}{}^{ab}(\w)R^{\m\n}{}^{ab}(\w)
    +\tfrac{1}{2}R_{\m\n}{}^{ab}(\w)\, {H}_{\r}{}^{ab}\, H^{\m\n\r}
\nonumber\\
&&\qquad\qquad\quad  +\tfrac{1}{8} (H^2)_{ab} (H^2)^{ab}
     + \tfrac{1}{24} H^4 \big\}
\label{L1BR2}\\
&&\qquad -  \tfrac{1}{2}\alpha \big\{
    R_\m{}^c(\w) H^{\m ab}H_{abc}
   + e^\m{}_c e^\n{}_d {\cal D}_\n H_{abd}{\cal D}_\m H_{abc}
\nonumber\\
&&\qquad\qquad\quad + 2 \partial_c\Phi\,H_{abd}{\cal D}_d H_{abc}
         - 2 \partial_d\Phi\,H_{abd}{\cal D}_c H_{abc}\big\} \big]\,.
\label{L1BRredef}
\eea
The term proportional to the Ricci tensor in 
(\ref{L1BRredef}) then 
contributes through a field redefinition 
to the terms quartic in $H$,
and gives an additional contribution 
involving derivatives of $\Phi$ (see (\ref{Ricciredef})). 
Using (\ref{eomB}) and partial integrations all remaining terms 
can be made to cancel.

The final result is then 
\bea
 {\cal L}_{\rm BR}&=&
 \tfrac{1}{2}e\,e^{-2\Phi}\big[ -R(\w) -
   \tfrac{1}{12} \bar{H}_{\m\n\r} \bar{H}^{\m\n\r} +
         4 \partial_\m\Phi\partial^\m\Phi 
\nonumber\\
&&\qquad  -\tfrac{1}{2}\alpha\,\big\{  
     R_{\m\n}{}^{ab}(\w)R^{\m\n}{}^{ab}(\w)
    +\tfrac{1}{2}R_{\m\n}{}^{ab}(\w)\, {H}_{\r}{}^{ab}\, H^{\m\n\r}
\nonumber\\
&&\qquad\qquad\quad  -\tfrac{1}{8} (H^2)_{ab} (H^2)^{ab}
    +\tfrac{1}{24} H^4 \big\}\big]\,,
\label{BRfinal}
\eea
in agreement with \cite{Metsaev:1987zx} if we set 
$R(\Gamma)=-R(\w)$ and $\alpha=-\tfrac{1}{4}\alpha'$, and adjust the 
overall normalisation. Of course \cite{Metsaev:1987zx} also includes
the LCS term in $H^2$ for the heterotic string effective action,
see the footnote in \cite{Metsaev:1987zx}, page 400.

\section{Higher orders and field redefinitions\label{High}}

In \cite{Bergshoeff:1989de} it was shown that the effective action
to order $\a^2$ consists of terms which are bilinear in the fermions
(\ref{L0BR},\ \ref{L1BR}). 
This is no longer true when the effective action at order $\a$ is in 
the form (\ref{BRfinal}).

Since the steps to go from (\ref{L0BR},\ \ref{L1BR}) to 
(\ref{BRfinal}) have all been explicitly determined, the effective action
at order $\alpha^2$ can in principle be constructed. Let us identify
the sources of bosonic ${\cal O}(\alpha^2)$-terms that we have encountered:
\begin{enumerate}
\item From the action (\ref{L0BR}) there are contributions outlined in
Appendix \ref{Expand}. We should now expand $\widetilde{H}$ to order $\a^2$,
which means that in ${\cal A}$ (\ref{calA}) also terms of order $\alpha$
should be considered. Then one should calculate $\widetilde{H}^2$.
\item $\bar{H}$ contains the LCS term of order $\alpha$. These
should now also be kept in the higher order contributions. 
\item In a number of places we have used the identity (\ref{DH}), the
resulting $R^2$ terms contribute to order $\alpha^2$.
\item We have used field redefinitions to modify the effective action
at order $\alpha$. A field redefinition is of the form 
\be
    e_\m{}^a \to e_\m{}^a + \alpha \Delta_\m^a\,,  
\ee
and is applied to the order $\alpha^0$ action. This has the effect of 
giving an extra contribution
\be
    \alpha \Delta_\m^a {\cal E}^\m{}_a
\ee
to the action, where ${\cal E}^\m{}_a$ is the Einstein equation at order
$\alpha^0$. Thus one can eliminate a term 
\be
   - \alpha \Delta_\m^a {\cal E}^\m{}_a\,.
\ee
at order $\alpha$. Contributions of order $\alpha^2$ arise 
because the transformation should also be applied to the order 
$\alpha$ action. 
\end{enumerate}

\noindent In this way the bosonic part of six-derivative terms in 
the effective action at order $\alpha^2$, 
corresponding to the order $\alpha$ action (\ref{L1MT}), can
be obtained, including the complete dependence on $H$. It would be interesting
to extend the calculation of black hole entropy of \cite{Sahoo:2006rp}
to this sector.

At order $\alpha^3$ the situation is different. In \cite{Bergshoeff:1989de}
an invariant related to the supersymmetrisation 
of the LCS terms was constructed. The status of $R^4$ invariants was
discussed in \cite{Tseytlin:1995bi}, with extensive reference to the
earlier literature. 

\section{Conclusions\label{Conc}}

We have established the equivalence between the effective actions of 
\cite{Bergshoeff:1989de} and \cite{Metsaev:1987zx} to order $\alpha$.
This indicates that the result of \cite{Sahoo:2006rp} might indeed
be a consequence of supersymmetry.

In principle the method of 
\cite{Sen:2005wa} can be extended to corrections with any number of 
derivatives. Supersymmetry provides the derivative contributions
at order $\ap{2}$, at $\ap{3}$ only partial results are known.
It would be interesting to extend the analysis of \cite{Sahoo:2006rp} to
include the next order. 

\acknowledgments

\bigskip

SP would like to thank Ashoke Sen for interesting and useful discussions.
SP thanks  the Centre for Theoretical Physics in Groningen for their
hospitality. WC and MdR  are supported by the European
Commission FP6 program MRTN-CT-2004-005104 in which WC and MdR
are associated to Utrecht University.

\appendix

\section{Calculational details\label{Conv}}
\subsection{Lagrangian density and redefinitions\label{LRsection}}

In \cite{Bergshoeff:1989de} the Lagrangian density 
takes on the form 
\be
  {\cal L}_{R} = \tfrac{1}{2}e\phi^{-3}\left(-R(\w) -
   \tfrac{3}{2} \widetilde{H}_{\m\n\r} \widetilde{H}^{\m\n\r} +
         9( \phi^{-1}\partial_\m\phi)^2\right)\,.
\label{LR1}
\ee
with the following definitions\footnote{In this letter we use
the the notation and conventions of \cite{Bergshoeff:1989de} and $\alpha$ is a
free parameter proportional to $\alpha'$, 
the inverse of the string tension.}:
\bea
   \widetilde{H}_{\m\n\r}&=&\partial_{[\m}B_{\n\r]}
      -\alpha\sqrt{2} \,{\cal
      O}_{3\,\m\n\r}(\W_{-})\,,\label{twoform}
\\
  {\cal O}_{3\,\m\n\r}(\W_-)&=&
      \W_{-[\m}{}^{ab}\partial_\n\W_{-\r]}{}^{ab}
    -\tfrac{2}{3}\W_{-[\m}{}^{ab}\W_{-\n}{}^{ac}\W_{-\r]}{}^{cb}\,,
\label{LCSA}
\\
  \W_{-\m}{}^{ab} &=& \w_\m{}^{ab} -\tfrac{3}{2}\sqrt{2}\widetilde{H}_{\m}{}^{ab}\,.
\eea 
Antisymmetrisation brackets are with weight 1.

First we redefine the fields to obtain agreement with the
conventions in \cite{Sahoo:2006rp}. The redefinitions are
\begin{enumerate}
\item The dilaton: the change is:
\be
    \phi^{-3}\to e^{-2\Phi}\,,\quad
    (\phi^{-1}\partial\phi)\to \tfrac{2}{3} \partial\Phi\,.
\ee
\item The 2- and 3-form fields: we set 
\be
 \widetilde{H}\to \tfrac{1}{3\sqrt{2}}\widetilde{ H}\,,\quad B\to \tfrac{1}{\sqrt{2}}B\,.
\ee
\end{enumerate}
${\cal L}_R$  now becomes 
\be 
{\cal L}_{R} =
\tfrac{1}{2}e\,e^{-2\Phi}\left(-R(\w) -
   \tfrac{1}{12} \widetilde{H}_{\m\n\r} \widetilde{H}^{\m\n\r} +
         4 \partial_\m\Phi\partial^\m\Phi\right)\,,
\label{LR} 
\ee 
as in (\ref{L0BR}).

The spin connection $\w(e)$ is the solution of 
\be
  {\cal D}_\m e_\n{}^a - {\cal D}_\n e_\m{}^a = 0\,,\ \quad
  {\rm with}\quad
  {\cal D}_\m e_\n{}^a \equiv \partial_\m e_\n{}^a
                     -\w_{\m}{}^{ac}e_{\n\,c}\,.
\label{dbeinpost} 
\ee 
The Riemann tensor and related quantities are defined as
\bea
   R_{\m\n}{}^{ab}(\w) &=& \partial_\m\w_\n{}^{ab}
                    -\partial_\n\w_\m{}^{ab}
                  -\w_\m{}^{ac}\w_{\n\,c}{}^b
                  +\w_\n{}^{ac}\w_{\m\,c}{}^b\,,
\\
  R_\m{}^a(\w) &=& e^{\n}{}_{b} R_{\m\n}{}^{ab}(\w) \,,
\\
  R(\w) &=& e^{\m}{}_a R_\m{}^a(\w)\,.
\eea 

\subsection{Equations of motion\label{eomL0}}

The equations of motion at order $\ap{0}$ are: 
\bea
 {\cal S} &=& ee^{-2\Phi} \big\{
          R(\w) 
      - 4 {\cal D}_a\partial^a\Phi 
      + 4 (\partial_a\phi)^2 
       + \tfrac{1}{12} H^{abc}H_{abc} \big\}
    \,,
\label{eomPhi}
\\
 {\cal B}^{\n\r} &=& \tfrac{1}{4}
    \partial_\m\big(ee^{-2\Phi}H^{\m\n\r}\big) = 0\,,
\label{eomB}
\\
 {\cal E}^\l{}_c &=& -\tfrac{1}{2}e^\l{}_c {\cal S}
  + e e^{-2\Phi}\,\big(
    R_c{}^\l(\w) + \tfrac{1}{4} (H^2)_\l{}^c
    -2e^\l_d {\cal D}_c\Phi\partial^d\Phi
    \big) = 0\,.
\label{eomE} 
\eea 
In the main text we use a field redefinition to
eliminate a contribution proportional to the Ricci tensor. The 
required equation is, modulo ${\cal E}$ and ${\cal S}$:
\bea
   R_\m{}^a(\w) = 2{\cal D}_\m\partial^a \Phi - 
          \tfrac{1}{4} (H^2)_\m{}^a \,. 
\label{Ricciredef}  
\eea

\subsection{Expanding ${\cal L}_R$ in powers of $\alpha$\label{Expand}}

The 3-form field $\widetilde{H}$ is defined recursively
by (\ref{Htilde},\ \ref{LCS},\ \ref{Om}). We find
\bea  
\widetilde{H}_{\m\n\r}&=&H_{\m\n\r}
      -6\alpha \,({\cal O}_{3\,\m\n\r}(w)+{\cal A}_{\m\n\r})
   =\bar{H}_{\m\n\r}-6\alpha{\cal A}_{\m\n\r} \, \,,
\label{H(W)}
\eea 
where ${\cal O}_{3\,\m\n\r}(w)$ is the gravitational contribution 
(order $\alpha^0$) of the Lorentz Chern-Simons term, and 
\bea
{\cal A}_{\m\n\r}&=&
 \tfrac{1}{2}\partial_{[\m}(\w_\n{}^{ab}\widetilde{H}_{\r]}{}^{ab})
-\tfrac{1}{2}R_{[\m\n}{}^{ab}(\w) \widetilde{H}_{\r]}{}^{ab}
+  \tfrac{1}{4}\widetilde{H}_{[\m}{}^{ab}
  {\cal D}_\n \widetilde{H}_{\r]}{}^{ab}
\nonumber\\
&&
+ \tfrac{1}{12}\widetilde{H}_{[\m}{}^{ab}\,\widetilde{H}_{\n}{}^{ac}
\,\widetilde{H}_{\r]}{}^{cb}\,.
\label{calA}
\eea 
To order $\alpha$ ${\cal L}_R$ (\ref{LR}) can be written  as
\bea
  {\cal L}_{R} &=&
 \tfrac{1}{2}e\,e^{-2\Phi}\big[-R(\w) -
   \tfrac{1}{12} \bar{H}_{\m\n\r} \bar{H}^{\m\n\r} +
         4 \partial_\m\Phi\partial^\m\Phi 
\nonumber\\
&& + \alpha\,\big\{
   \tfrac{1}{2} {H}^{\m\n\r}\partial_{\m}(\w_\n{}^{ab}{H}_{\r}{}^{ab})
  -\tfrac{1}{2} R_{\m\n}{}^{ab}(\w)\, {H}_{\r}{}^{ab}\, {H}^{\m\n\r}
+  \tfrac{1}{4} {H}^{\m\n\r}{H}_{\m}{}^{ab}
                {\cal D}_\n {H}_{\r}{}^{ab}
  +\nonumber\\
&&\qquad
 + \tfrac{1}{12} {H}^{\m\n\r}\,{H}_{\m}{}^{ab}\,
                 {H}_{\n}{}^{ac}{H}_{\r}{}^{cb} \big\}\big]\,.
\label{LRfinal}
\eea
The term with the $H\partial(\w H)$ is, after partial integration,
proportional to (\ref{eomB}) and can be eliminated by a field 
redefinition.

\subsection{Simplification of ${\cal L}_{R^{2}}$ terms \label{termLR2}}

We often use the identity 
\be
 {\cal D}_{[a}(\W_-)\widetilde{H}_{bcd]} =
     -\tfrac{3}{2}\alpha\,R_{[ab}{}^{ef}(\W_-) R_{cd]}{}^{ef}(\W_-)\,,
\label{DH}
\ee 
to isolate terms that are of higher order in $\a$.
The term (\ref{term4}) can be simplified by using the cyclic
identity for the Riemann tensor:
\be
   R_{\m\n}{}^{ab}(\w) H^{\m ac}H^{\n cb} =
   -\tfrac{1}{2} R_{\m\n}{}^{ab} H^{\m\n c}H^{abc}\,.
\label{RHH} 
\ee 
Now we consider (\ref{term5}). Note
that the two terms written in (\ref{term5}) are in fact the same.
Then we have 
\bea 
\tfrac{1}{2}
  \big({\cal D}_\m H_\n{}^{ab}-
       {\cal D}_\n H_\m{}^{ab}\big)H^{\m ac}H^{\n cb} &=&
-   {\cal D}_\m H_\n{}^{ab} H^{\m ac}H^{\n bc} 
\nonumber\\
&=& - {\cal D}_{[e} H_{fab]} \,H^{eac}H^{fbc} \,.
\label{HHDH} 
\eea 
This
term is completely of order $\ap{2}$.
Finally we consider (\ref{term3}). This can be expressed as
\bea
&& \tfrac{1}{2} e e^{-2\Phi}
   \big({\cal D}_\m H_\n{}^{ab}-
       {\cal D}_\n H_\m{}^{ab}\big)
     {\cal D}^\m H^{\n ab} = ee^{-2\Phi}\big( 
        2 R_{\m\n}{}^{ab}H^{\m ac}H^{\n\,cb}
                     + R_\m{}^c H^{\m ab}H_{abc}
\nonumber\\
&&\quad
  + e^\m{}_c e^\n{}_d {\cal D}_\n H_{abd}{\cal D}_\m H_{abc}
 + 2 \partial_c\Phi \,
  H_{abd}{\cal D}_d H_{abc}
  - 2 \partial_d\Phi\,
  H_{abd}{\cal D}_c H_{abc}
\nonumber\\
&&\quad
 + 2 {\cal D}_c H_{abd} {\cal D}_{[c} H_{abd]} \big) \,.
\label{DHDH}
\eea 
The last term is of order $\alpha'^2$.


\begin{thebibliography}{mmm}

\bibitem{Mohaupt:2000mj}
  T.~Mohaupt,
  ``Black hole entropy, special geometry and strings,''
  Fortsch.\ Phys.\  {\bf 49}, 3 (2001)
  [arXiv:hep-th/0007195].
\bibitem{Gross:1986mw}
  D.~J.~Gross and J.~H.~Sloan,
  ``The Quartic Effective Action for the Heterotic String,''
  Nucl.\ Phys.\  B {\bf 291} (1987) 41.
\bibitem{Metsaev:1987zx}
  R.~R.~Metsaev and A.~A.~Tseytlin,
  ``Order alpha-prime (Two Loop) Equivalence of the String Equations of Motion
  and the Sigma Model Weyl Invariance Conditions: Dependence on the Dilaton and
  the Antisymmetric Tensor,''
  Nucl.\ Phys.\  B {\bf 293} (1987) 385.
\bibitem{Hull:1987pc}
  C.~M.~Hull and P.~K.~Townsend,
  Phys.\ Lett.\  B {\bf 191} (1987) 115.
\bibitem{Bergshoeff:1988nn}
  E.~Bergshoeff and M.~de Roo,
  ``Supersymmetric Chern-Simons terms in ten-dimensions,''
  Phys.\ Lett.\  B {\bf 218} (1989) 210.
\bibitem{Bergshoeff:1989de}
  E.~A.~Bergshoeff and M.~de Roo,
  ``The Quartic Effective Action Of The Heterotic String And Supersymmetry,''
  Nucl.\ Phys.\  B {\bf 328} (1989) 439.
\bibitem{Sahoo:2006rp}
  B.~Sahoo and A.~Sen,
  JHEP {\bf 0609} (2006) 029
  [arXiv:hep-th/0603149].
\bibitem{Sen:2005wa}
  A.~Sen,
  ``Black hole entropy function and the attractor mechanism in higher
  derivative gravity,''
  JHEP {\bf 0509} (2005) 038
  [arXiv:hep-th/0506177].
\bibitem{Tseytlin:1995bi}
  A.~A.~Tseytlin,
  ``Heterotic - type I superstring duality and low-energy effective actions,''
  Nucl.\ Phys.\  B {\bf 467} (1996) 383
  [arXiv:hep-th/9512081].

\end{thebibliography}
\end{document}